\documentclass[10pt]{article}

\usepackage{verbatim}
\usepackage[numbers]{natbib}
\usepackage{graphicx}
\usepackage{url}
 
\begin{document}

\title{One- and two-sample nonparametric tests for the signal-to-noise ratio based on record statistics}

\author{
Damien Challet$^{1,2}$\\
$^{1}$Laboratoire de math\'ematiques appliqu\'ees aux syst\`emes,\\ CentraleSup\'elec, \\92295 Ch\^atenay-Malabry CEDEX,
France\\
$^{2}$Encelade Capital SA, EPFL Innovation Park, \\1015 Lausanne, Switzerland\\
}

\maketitle


\begin{abstract}
A new family of nonparametric statistics, the r-statistics, is introduced. It consists of counting the number of records of the cumulative sum of the sample. The single-sample r-statistic is almost as powerful as Student's t-statistic for Gaussian and uniformly distributed variables, and more powerful than the sign and Wilcoxon signed-rank statistics as long as the data are not too heavy-tailed. 

Three two-sample parametric r-statistics are proposed, one with a higher specificity but a smaller sensitivity than Mann-Whitney U-test and the other one a higher sensitivity but a smaller specificity. A nonparametric two-sample r-statistic is introduced, whose power is very close to that of Welch statistic for Gaussian or uniformly distributed variables.
\end{abstract}
~\\
keywords: {nonparametric statistics, signal-to-noise ratio, statistical power, AUC, record statistics}


\section{Introduction}

Nonparametric statistics play a special role in data analysis as they are usually more robust and require less assumptions about the underlying data distribution \cite{gibson1976nonparametric}. Well-known nonparametric statistics, such as sign and Wilcoxon signed-rank for single samples, and Mann-Whitney U-statistic for two samples are however much less powerful than the parametric t- or Welch statistics for Gaussian or uniformly distributed variables, while the opposite holds for fat-tailed data. Here I propose a new type of nonparametric statistics, called r-statistics, which is almost as powerful as t- and Welch statistics for Gaussian variables and better than the all the above for not too fat-tailed variables. As a consequence, they provide a robust alternative to usual statistics.



\maketitle

Let us write down the definition of the t-statistic as a way to introduce useful notations.
Take a sample of $N$ values of a quantity of interest, denoted by $\{x_{n}\}$,
$n=1,\cdots,N$, assumed to be independently identically distributed
(iid). Denoting an estimate with a hat, the t-statistic of the sample
is $\hat{t}=\hat{\theta}\sqrt{N}$ where $\hat{\theta}=\hat{\mu}/\hat{\sigma}$
is its estimated signal-to-noise ratio (SNR thereafter), $\hat{\mu}$
its estimated average and $\hat{\sigma}$ its estimated standard deviation.

The robustness of commonly used nonparametric statistics is due in part to the fact that they reduce sample values to integer quantities, such as ranks and signs, from which the statistics are computed. The same recipe underlies r-statistics which 
 are based on the (integer) number of records of the cumulative sum (or equivalently the integrated signal) of the sample values defined as  $\xi_N=\{X_t\}_{1\le t\le N}$ where $X_{t}=\sum_{n=1}^{t}x_{n}$, $1\le t\le N$. If the distribution of $x$ has a zero average, $X_{t}$ is nothing else than the position of an unbiased random walker at time $t$. 
 A remarkable result, based on Sparre Andersen theorem \cite{andersen1953fluctuations}, states 
that the distribution of the number of upper records (or equivalently the number of jumps of the running maximum) in $N$ steps,
denoted by $R_{+}$, does not depend on the distribution
of $x_{n}$ as long as it is symmetric (i.e. $x$ and $-x$ are equiprobable)
and continuous, and the sample values are uncorrelated \cite{majumdar2008universal}; note that the first point is always considered as the first upper (and lower) record (see Fig. \ref{fig:permutations}). In addition, this distribution is known exactly:
\begin{equation}
P(R_{+},N)={2N-R_{+}+1 \choose N}/2^{2N-R_{+}+1},\label{eq:PRN}
\end{equation}
which tends to a Gaussian distribution $\mathcal{N}(\sqrt{4N/\pi},{(4-2/\pi)N})$
for large $N$ \cite{majumdar2008universal}. For symmetry reasons,
the number of lower records (i.e., the number of jumps of the running minimum), denoted by $R_{-}$, follows the same distribution.  This result has spawned many studies on so-called record statistics (see \cite{majumdar2012record} for a review). 

\section{Single-sample statistics}

Even if single-sample statistic is a well-trodden domain in statistics, using an ever so slightly more powerful statistic provides an invaluable advantage in competitive situations such as speculative trading or friend-or-foe identification. One of the problems of single-sample nonparametric statistics is that they are less powerful than a t-statistic for Gaussian or uniformly-distributed variables. The r-statistic  remedies this problem while being robust. Note that Sparre Andersen's assumption of symmetric distribution is shared by Wilcoxon signed-rank statistic.

\begin{figure*}
\centerline{\includegraphics[width=0.9\textwidth]{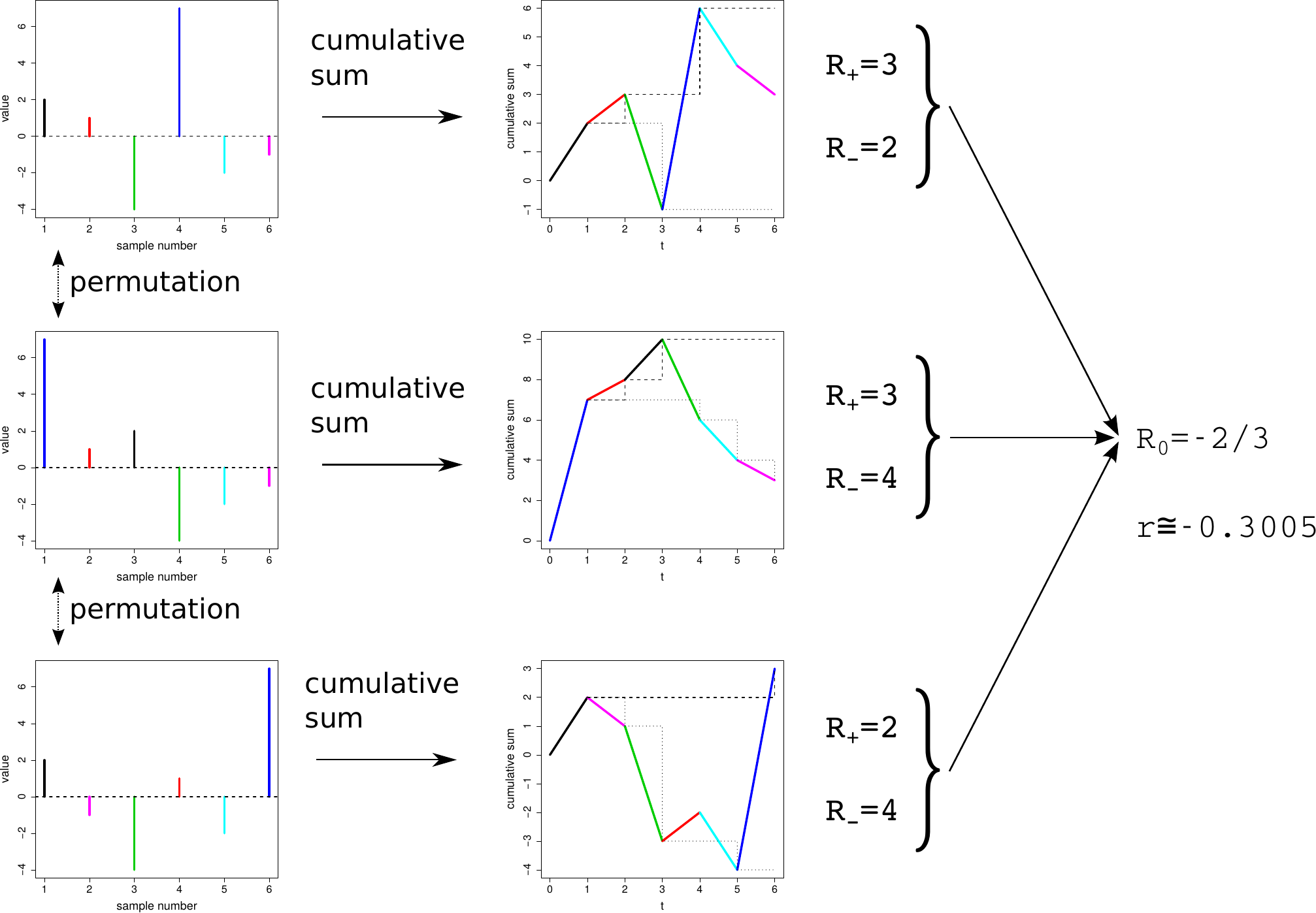}}
\caption{Schematic explanation of the idea behind the r-statistics: one computes the difference between the number of jumps of the running maximum (dashed lines) and the number of jumps of the running minimum (dotted lines) of the cumulated sums of the sample values, averaged over many random permutations. By convention, the first point counts as a first jump for both the running maximum and minimum. The r-statistic $r\simeq 0.3005$ is simply $R_0/\sigma_N$ where $\sigma_N\simeq 1.97$ for $N=6$ (see Eq.\ (\ref{eq:rstat})).\label{fig:permutations}}
\end{figure*}

Up to this point, the quantities $R_+$ and $R_-$  have two flaws as statistics: first, they are bounded from below by zero, thus it is much easier to design a statistical test based on their difference $R_0=R_{+}-R_{-}$. The number of upper records of $X_N$ has a straightforward interpretation: $R_+$ is nothing else than the number of time steps during which $X_{N}$ is not in a drawdown (i.e., not below its running maximum). Thus, the quantity $R_+-R_-$ is the time spent in a drawup minus the time spent in a drawdown.

Second, $R_0$ is by definition an integer number, which may be detrimental to both statistical power and efficiency. The key new    
idea is to note that, for iid data $x_{n}$, the integrated signal
of any random permutation of $\{x_{n}\}$ is as valid a representation as
$X_{N}$. Thus one can compute the average number of records of
$R_{0}$ over $P\gg1$ random permutations, denoted by $\bar{R}_{0}$. Figure \ref{fig:permutations} explains this idea graphically.

Let us simply write $R_0$ instead of $\bar{R_0}$ in the following for the sake of readability. By definition, the distribution of $R_0$ converges towards a Gaussian distribution of zero average. Because the numbers of upper and lower records of a given random walk are correlated in an unknown way, the standard deviation of the distribution of $\bar{R}_0$, denoted by $\sigma_N$, must be measured numerically for the time being. Extensive numerical simulations (see Appendix A) show that $\sigma_N=1.66(1-0.88N^{-1/2})\sqrt{(2-4/\pi)N}$, thus the single-sample r-statistic is defined as

\begin{equation}\label{eq:rstat}
r=R_0\frac{1}{1.66(1-0.88N^{-1/2})\sqrt{(2-4/\pi)N}}.
\end{equation}
Asymptotically $P(r)\to \mathcal{N}[0,(\sigma_N)^2]$, but the convergence to Gaussian distribution is quite slow. For example, $P(R_0)$ is Gaussian up to 2 standard deviations for $N=1000$ (see Appendix A); thus, for the time being, to build a statistical test  one must resort to estimating the distribution of $P(R_0)$ numerically for a given $N$ and use it to obtain p-values. Computations are quick (and the full source code is available). 

\begin{figure*}
\centerline{\includegraphics[width=0.5\textwidth]{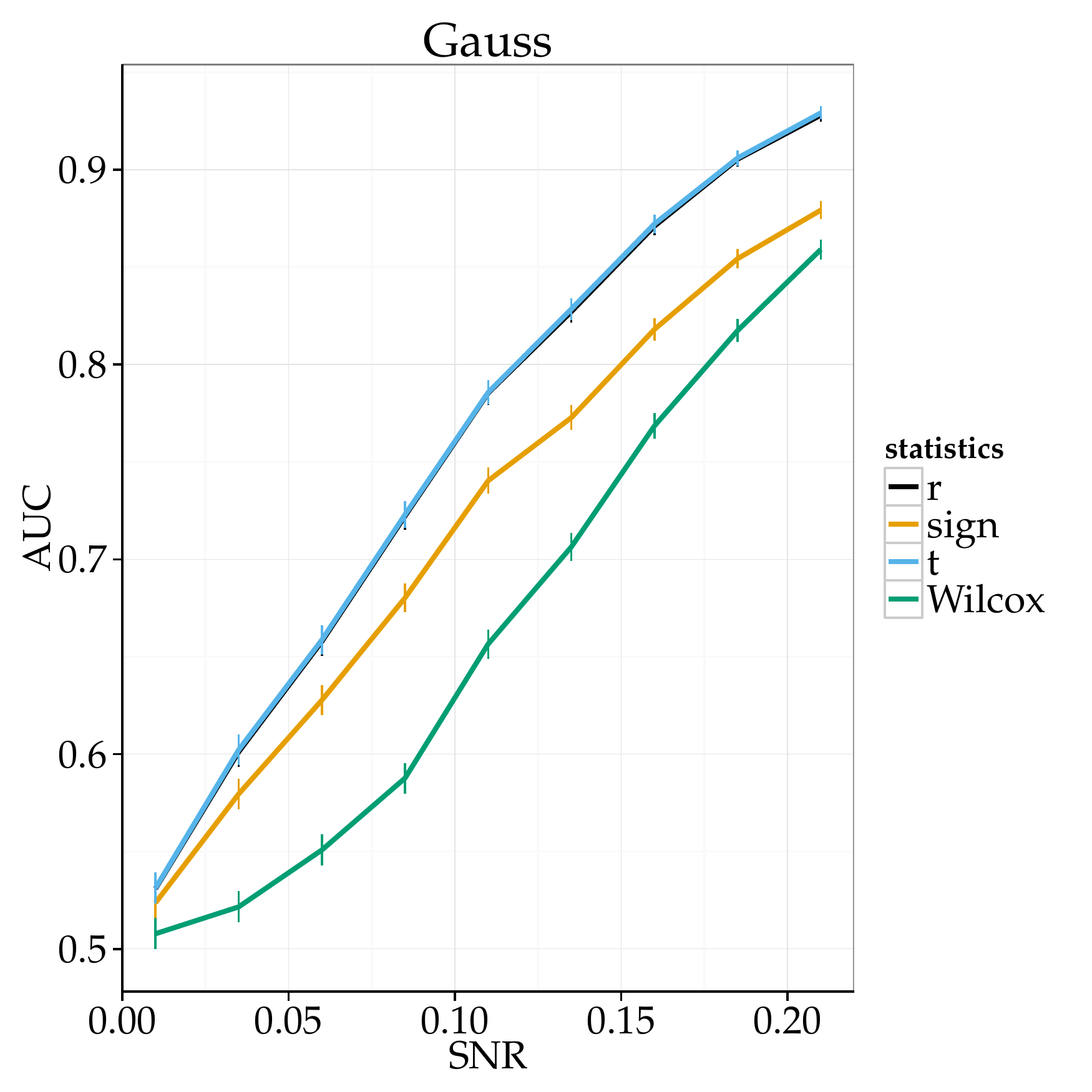}\includegraphics[width=0.5\textwidth]{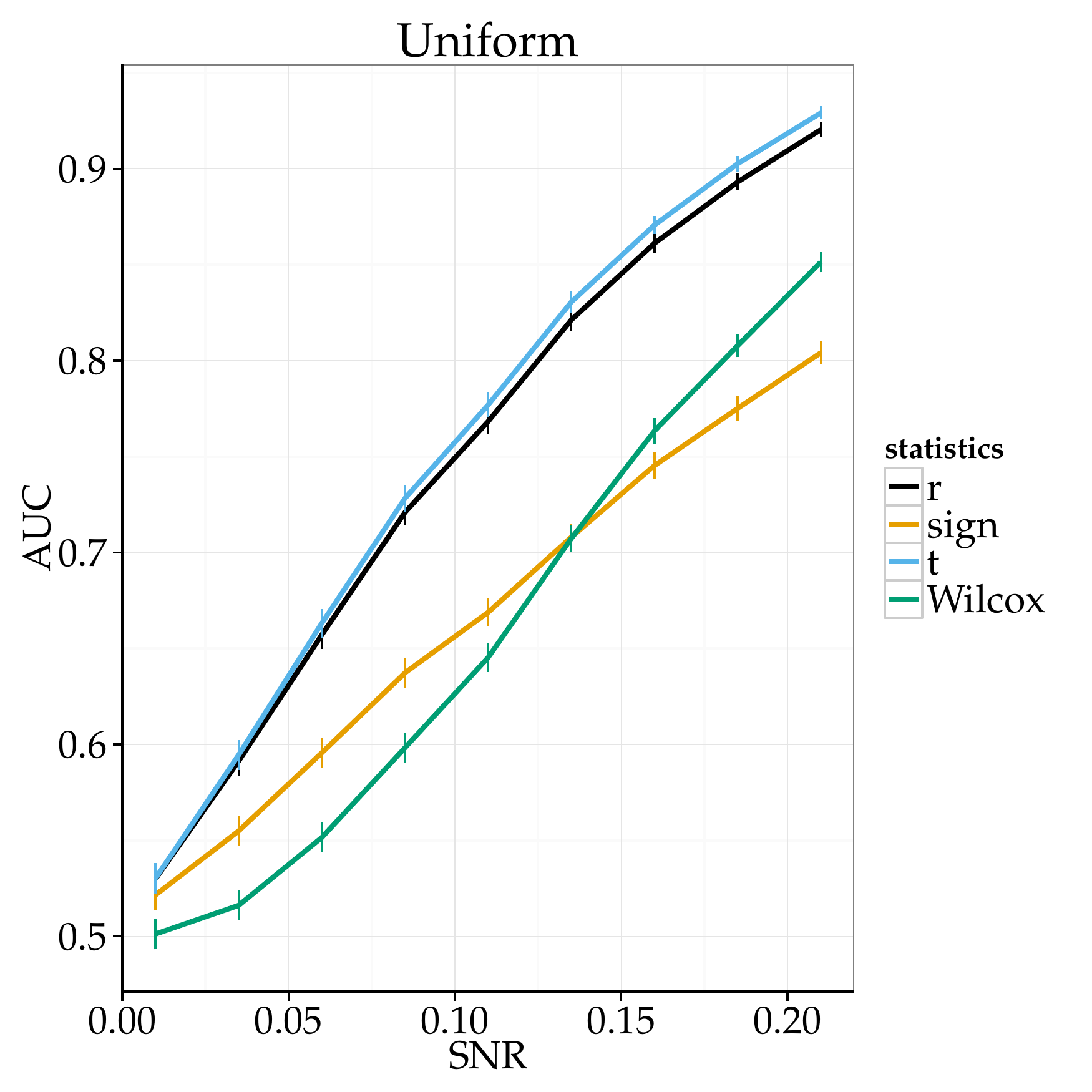}}

\caption{Area under curve (AUC) versus the signal-to-noise ratio $\theta=\mu/\sigma$ of the alternative; $N=100$, 
10000 samples per point, 10000 random permutations per sample. Error bars
set at two standard deviations. Continuous
lines are only meant for eye-guidance.\label{fig:AUC_vs_SNR}}
\end{figure*}

\begin{figure*}
\centerline{\includegraphics[width=0.5	\textwidth]{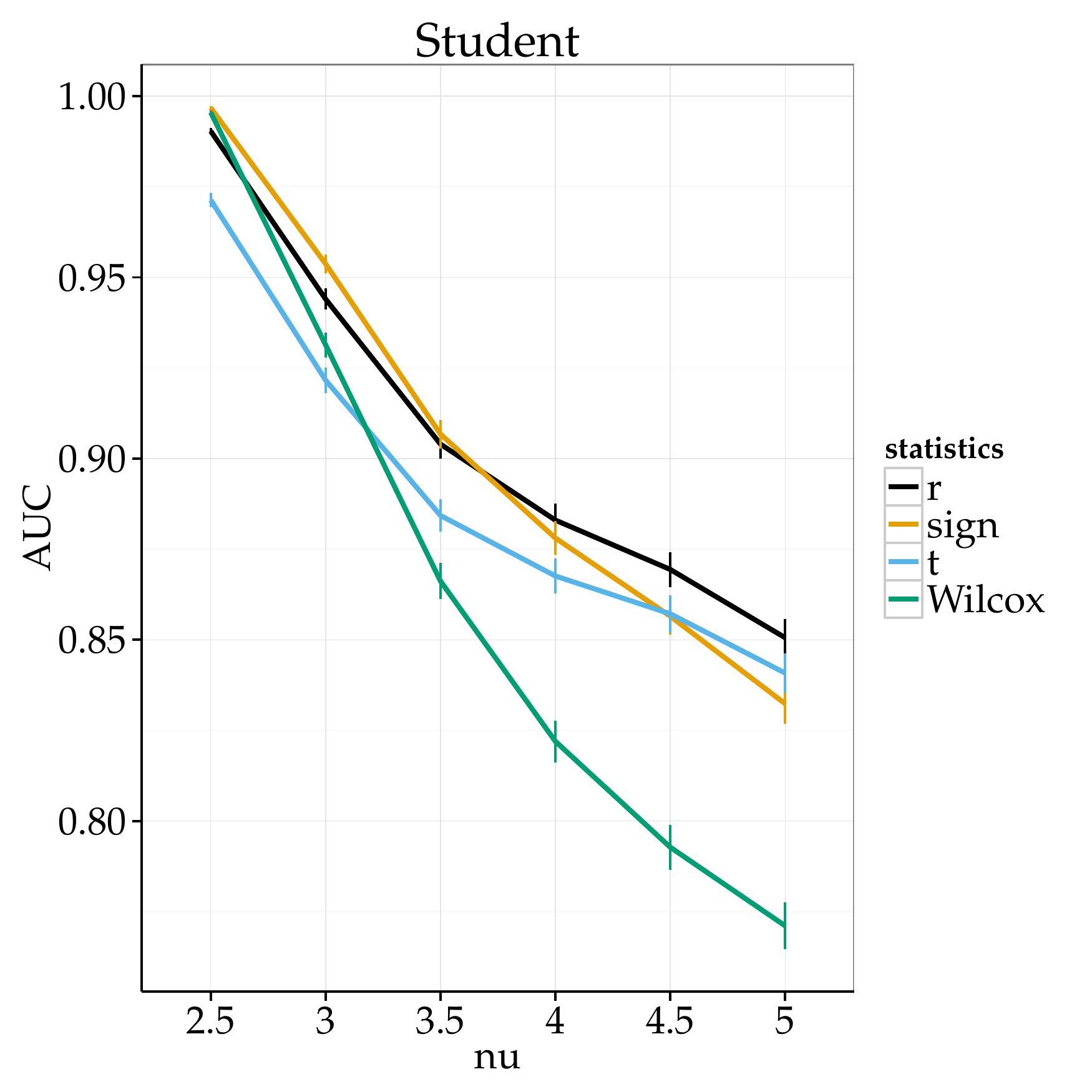}\includegraphics[width=0.5	\textwidth]{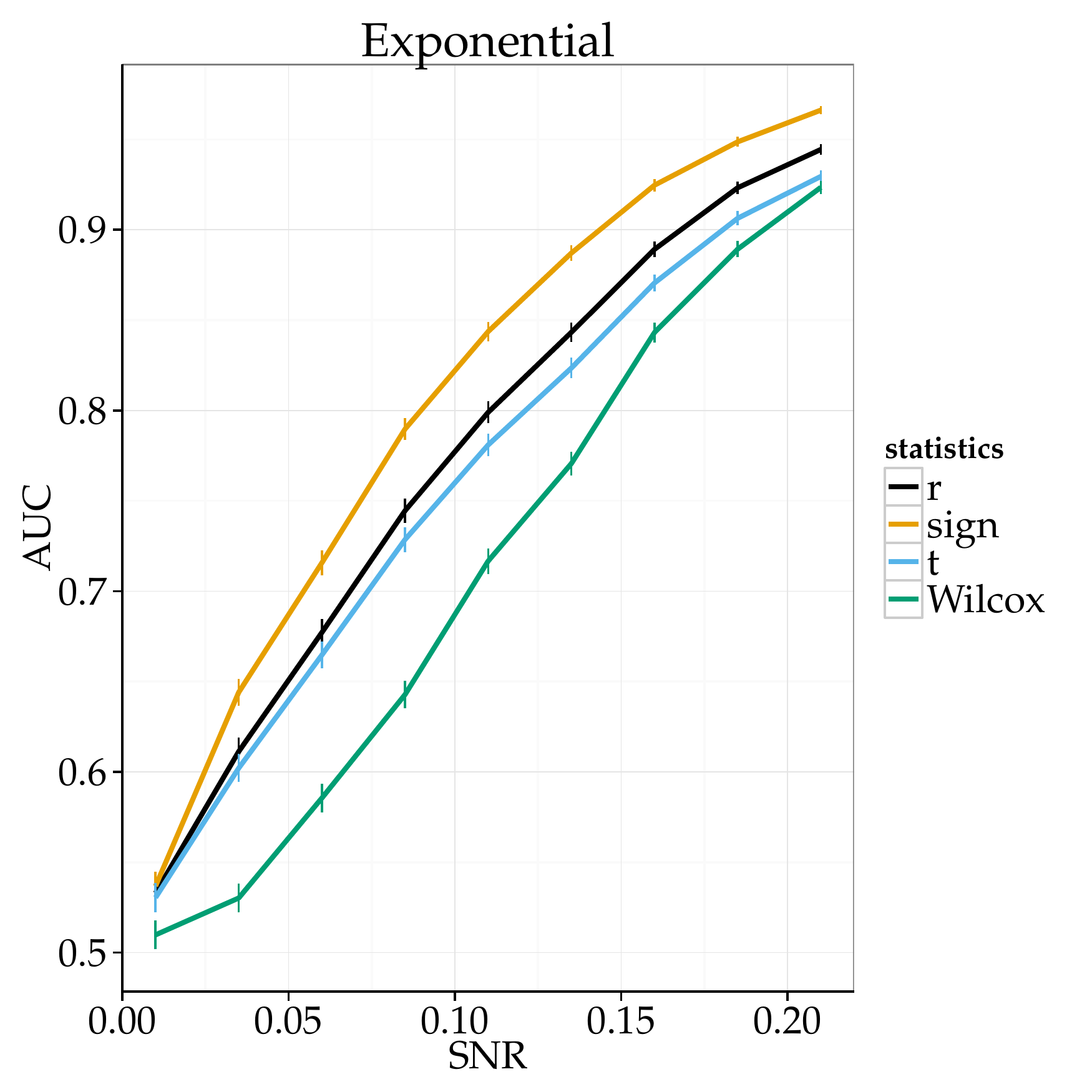}}

\caption{Area under curve (AUC) vs the signal-to-noise ratio $\theta=\mu/\sigma$
for various types of distributions of $\{x_{n}\}$. $N=100$, 10000
samples per point, 10000 random permutations per sample. Error bars set at
two standard deviations.
 Continuous lines are
only meant for eye-guidance.\label{fig:AUC_vs_snr_heavy}}
\end{figure*}

Assessing the power of the single-sample r-statistic requires to estimate $P(R_0)$ for $\theta=0$ and for the alternative $\theta\ne0$ (separately), and then to compute the Receiver Operating Characteristic (ROC) curve of the r-statistic \cite{hastie2009elements}. ROC curves for r-, t-, sign, and Wilcoxon signed-rank sum statistics are reported in Appendix \ref{sec:ROC_curves}. The ROC curves of r-statistics do not cross those of the other statistics, hence the Area Under Curve (AUC), a scalar summary of statistical power measured in ROC curves (the larger, the
better), is meaningful for comparing the power of r-statistic with that of other statistics. Let us start with Gaussian variables. T-statistic is uniformly most powerful in this case \cite{neyman1933problems}, hence one expects that its AUC is the largest of all. Figure \ref{fig:AUC_vs_SNR} shows that  while sign and Wilcoxon statistics are much less powerful than t-statistic for Gaussian variables (as it is well known), r-statistic has very nearly the same power as t-statistics.  Uniformly distributed variables lead to  similar results (same figure). Generically, the relative power of r-statistic with respect to that of sign and Wilcoxon statistics decreases as the tails of the data become heavier. This is illustrated in Fig.\ \ref{fig:AUC_vs_snr_heavy} which reports the AUC versus the $\nu$, the tail parameter of Student's t-distribution (used as a parametric way to obtain heavy-tailed data). Wilcoxon statistic becomes more powerful than r-statistic for $\nu\simeq 2.5$, while sign statistic wins when $\nu<3.5$. The same behaviour is found for exponentially distributed variables (same figure), in which case sign statistic is better than r-statistic.

One of the assumptions of the r-statistic is that the increments have a zero average, but this does not tell what the alternative is. When the average increment is not zero, but still comes from a symmetric distribution around its average, the average record number of such random walks is a function of the signal-to-noise ratio $\theta=\mu/\sigma$ \cite{wergen2011record,wergen2012record,majumdar2012record}. Thus the r-statistic is a test for the signal-to-noise ratio.

\section{Two-sample situation} 

Building a two-sample version r-statistic may be done in several ways. Let us denote the two samples by $x=\{x_n\}$, $n=1,\cdots,N_x$, and $y=\{y_m\}$, $m=1,\cdots,N_y$. Assuming for the time being that $N_x=N_y$,  the simplest idea is to test if the difference of sample elements. If the two samples are paired, then $z=\{z_n=x_n-y_n\}$, i.e., the same random permutation must be applied to both sample elements; otherwise, independent permutations may be applied to $x$ and $y$. the r has zero signal-to-noise ratio, hence average, which amounts to computing the r-statistic of $\{z\}$, denoted $R_z$ in the following. It is nonparametric by definition.

Note that if the two samples are paired, then one should 

Another approach is to compute record statistics for each sample and then compare them. For example, one can use the difference between the number of upper (or lower) records of both samples, i.e., 
\begin{eqnarray}
\bar{R}_{+}^{(2)}&=&\bar{R}_+(x)-\bar{R}_+(y)\\
\bar{R}_{-}^{(2)}&=&\bar{R}_-(x)-\bar{R}_-(y).
\end{eqnarray}
This suggests a fourth statistics, $R_d=\bar{R}_{+}^{(2)}-\bar{R}_{-}^{(2)}$. Given the fact that the expected number of records associated with a sample of non-zero average is a function of both the signal-to-noise ratio and of the distribution of the sample values, the distributions of these three statistics have zero average if both samples have the same distribution and the same signal-to-noise ratio, which is therefore their associated null hypothesis. These three statistics would be nonparametric if their standard deviation was nonparametric. This is not the case, as shown by Ref.\ \cite{majumdar2012record} which gives a generic expression of the distribution-dependent prefactor of this quantity. Hence, $R_{\pm}^{(2)}$ and $R_d$ are bound to be parametric.

\begin{figure*}
\centerline{\includegraphics[width=0.5	\textwidth]{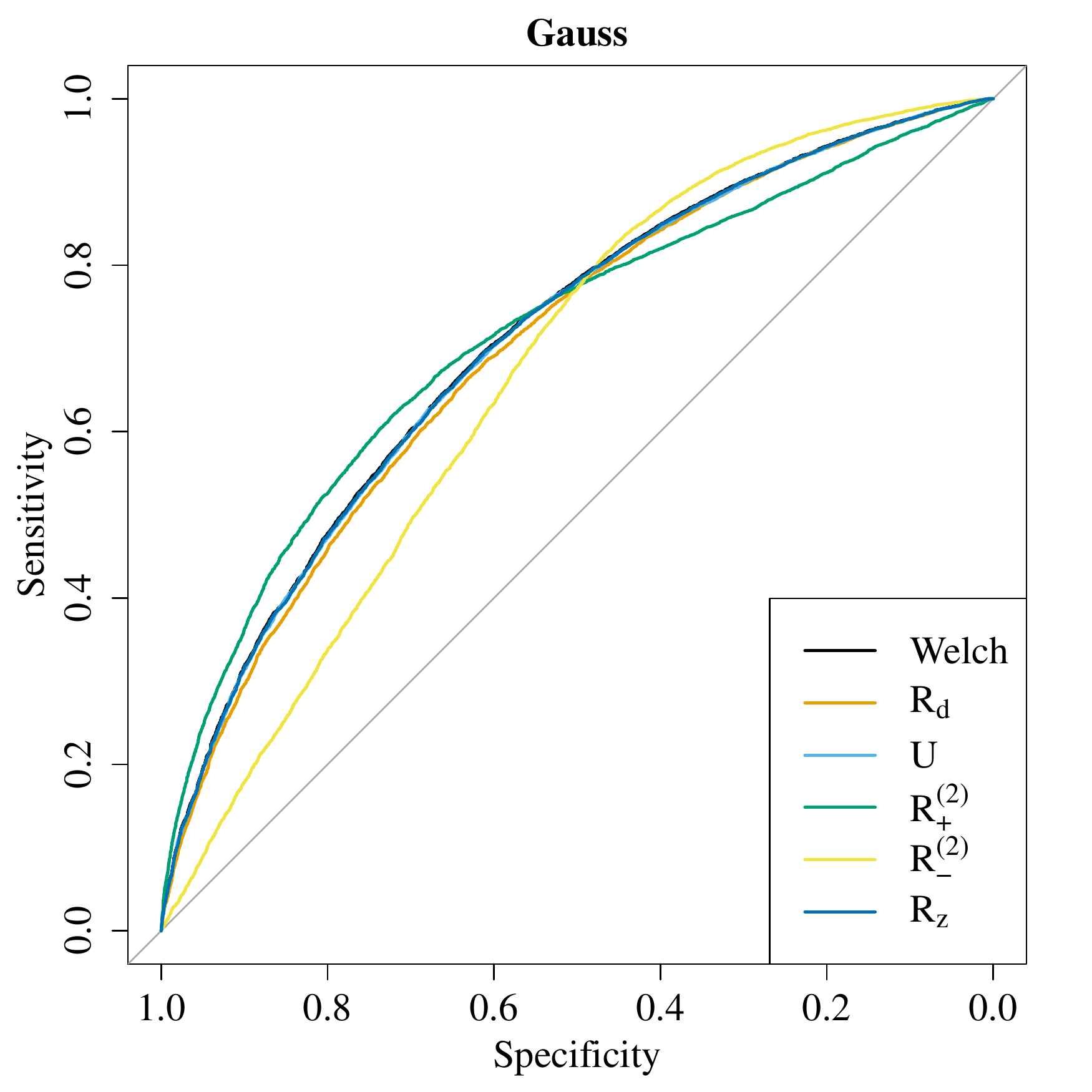}\includegraphics[width=0.5	\textwidth]{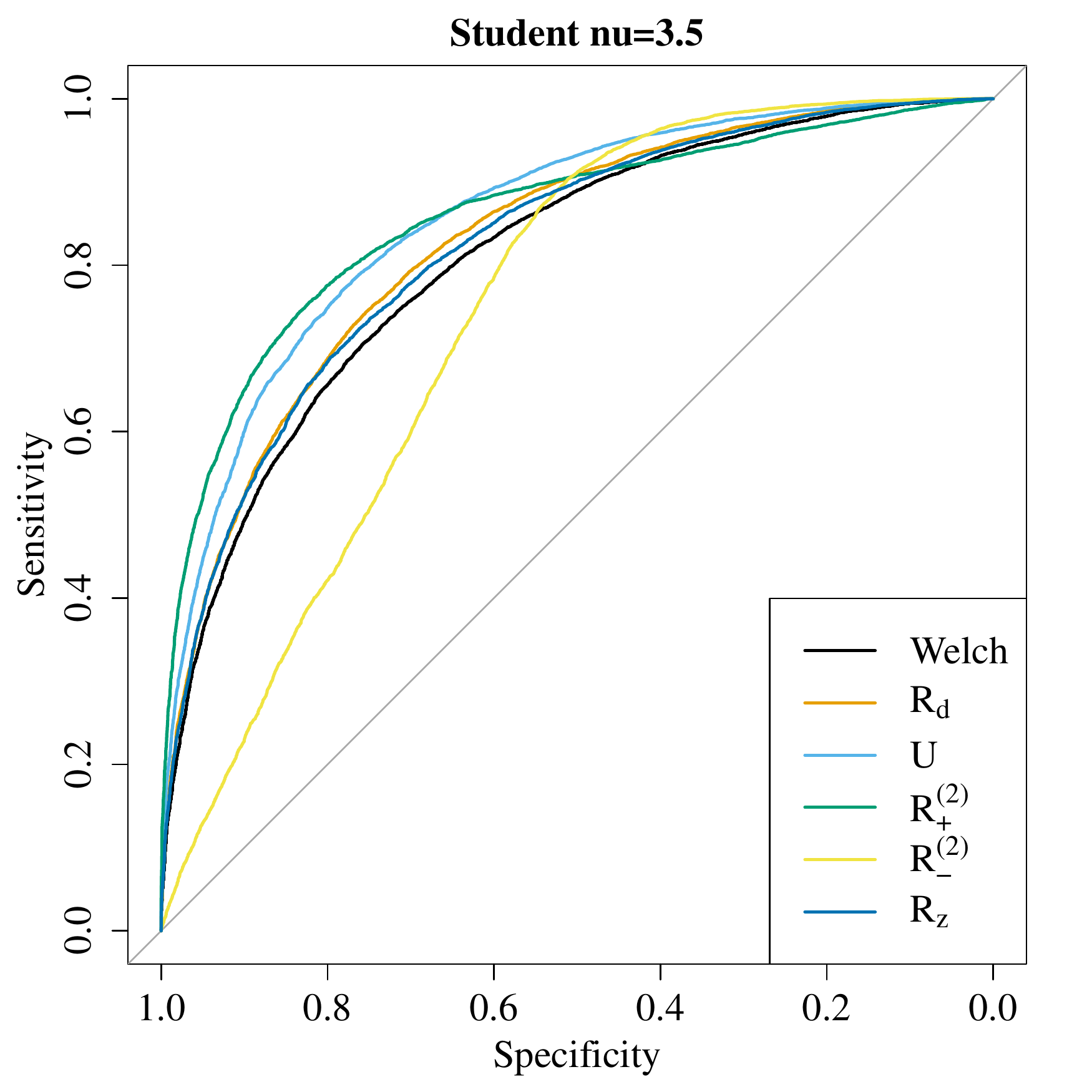}}

\caption{Two-sample situation: ROC curves for Gaussian (left plot) and Student t-distributed variables with $\nu=3.5$ (right plot) for the four  r-statistics, Mann-Whitney U-statistic and Welch statistic. Both samples have $\sigma=1$, while $E(x)=0$ and $E(y)=1$. $N=100$, 10000
samples per point, 10000 random permutations per sample. Specificity is equal to 1-false positive rate, and sensitivity is the true positive rate.l  \label{fig:ROC_2sample}}
\end{figure*}

Figure \ref{fig:ROC_2sample} shows ROC curves for $R_{\pm}^{(2)}$ and $R_d$ statistics when both samples have the same distribution, the same length and one of them has zero average. ROC curves for all distributions have common characteristics: generically,  $R_{+}^{(2)}$ has the largest specificity in the limit of large specificity and smallest sensitivity in the limit of large sensitivity of all statistics tested here, while the ${R}_{-}^{(2)}$ is the exact opposite. $R_d$ and $R_z$ have approximately the same power as a Welch statistic for distributions with mild tails (Gaussian and uniform), but do worse than Mann-Whitney otherwise, thus, preferring $R_z$ over $R_d$ makes sense.

There are two subtleties that apply to all two-sample versions of the r-statistics: first, The second subtlety arises when the two samples do not have the same number of elements, denoted   by $N_x$ and $N_y$ respectively. One solution consists in computing the record statistics of permutations of $\min(N_x,N_y)$ elements from each sample.  Because of the random nature of the permutations, this scheme ensures a fair sampling of the larger sample; another possibility is to keep all elements of the larger sample and to resample the smaller sample so as to have two samples of equal length.

\section{Conclusion}

While the r-statistics already have some direct applications, three important cases still need to be investigated. First, r-statistics as introduced here are only valid for uncorrelated data. While there is no exact result about record statistics of correlated random walks, numerical simulations point to simple corrections in specific cases \cite{wergen2014modelingrecordstock}; whether r-statistics remain nonparametric for non-iid variables remains to be tested. Practically, a simple modification of the way in which r-statistics are computed  respects short-range correlation:  one should permute
blocks of data instead of single values, much like block bootstraps
(e.g. \cite{carlstein1986use,kunsch1989jackknife,liu1992moving}); the length of the blocks may be found in a self-consistent way \cite{ledoit2008robust}. The second case is discrete distributions. While Sparre Andersen theorem is only valid for continuous variables, record statistics of random walks with discrete increments is similar \cite{majumdar2008universal}.
Finally, the case of non-symmetric distributions, e.g., the role of the skew is to be investigated. 

The fact that the r-statistics reflect signal-to-noise ratios is of particular interest to finance, because the performance of two assets (or trading strategies) are traditionally assessed with their Sharpe ratio, which is conceptually a signal-to-noise ratio. The usual methods are essentially equivalent to a Welch-statistic of their difference, computed with more sophisticated methods such as bootstraps and generalized moment methods \cite{lo2002sharpe,ledoit2008robust}. The latter require the estimation of first and second moments of the samples; some of them require the third and the fourth moments, which is problematic for asset prices \cite{lo2002sharpe,mertens2002lo,christie2005sharpe}. Further work explores the efficiency of r-statistic as a signal-to-noise estimator \cite{challet2014studentrecords}. 

As a final note, the universality of record statistics of unbiased random walks with symmetric increments extends to the time between two records \cite{majumdar2008universal}, which however does not yield a more powerful statistics.

The author thanks Gilles Fa\"y for his suggestions.

Full source code (R and C++) available  at \url{https://github.com/damienchallet/rstatistics}.

\appendix
\section{Asymptotic behaviour of the single-sample $R_0$}

\subsection{Standard deviation}
\begin{figure}
\centerline{\includegraphics[width=0.6\textwidth]{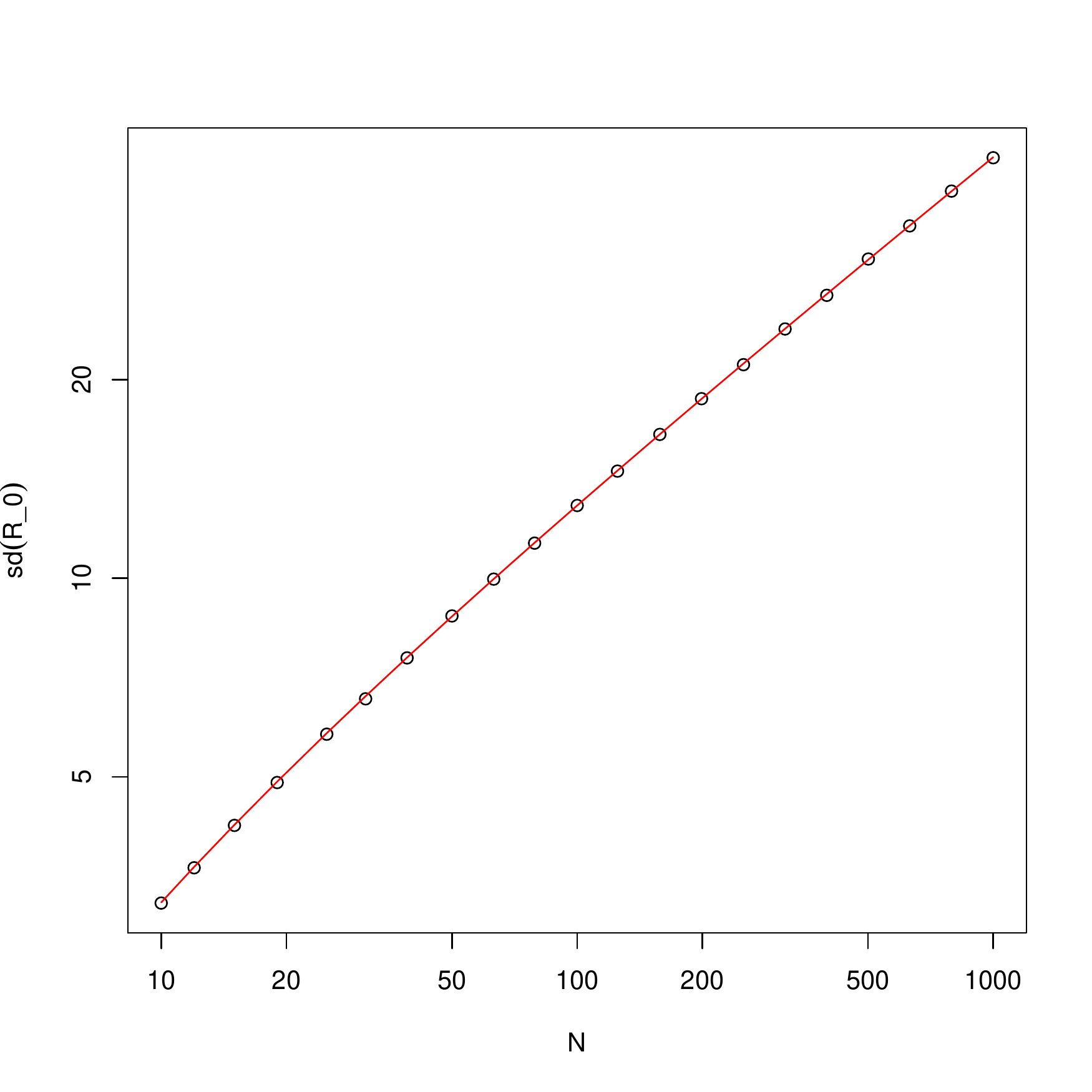}}
\caption{Standard deviation of $R_0$ as a function of $N$: numerical simulations (circles) and non-linear fit (continuous line) \label{fig:sdR}; $10^5$ samples perl point of  $R_0$ were computed for Gaussian variables, with $10^5$ permutations for each sample. }
\end{figure}

Numerical simulations were performed for $N=\texttt{floor}(10*(100^{1/20})^k)$ for $k=0,1,\cdots,20$, thus $N\in[10,1000]$: $10^5$ samples of  $R_0$ were computed for Gaussian variables ($10000$ random permutations for each sample). Then a non-linear fit
\begin{equation}
\sigma_N=\sqrt{(2-4/\pi)N}[a(1-bN^{-c})]\label{eq:sdR}
\end{equation} 
yields $a=1.659\pm0.008$, $b=0.88\pm0.04$, $c=0.5\pm0.02$  (errors set at two standard deviations); the goodness of fit is obvious in Figure \ref{fig:sdR}.

\subsection{Convergence to a Gaussian distribution}

\begin{figure}
\centerline{\includegraphics[width=0.6\textwidth]{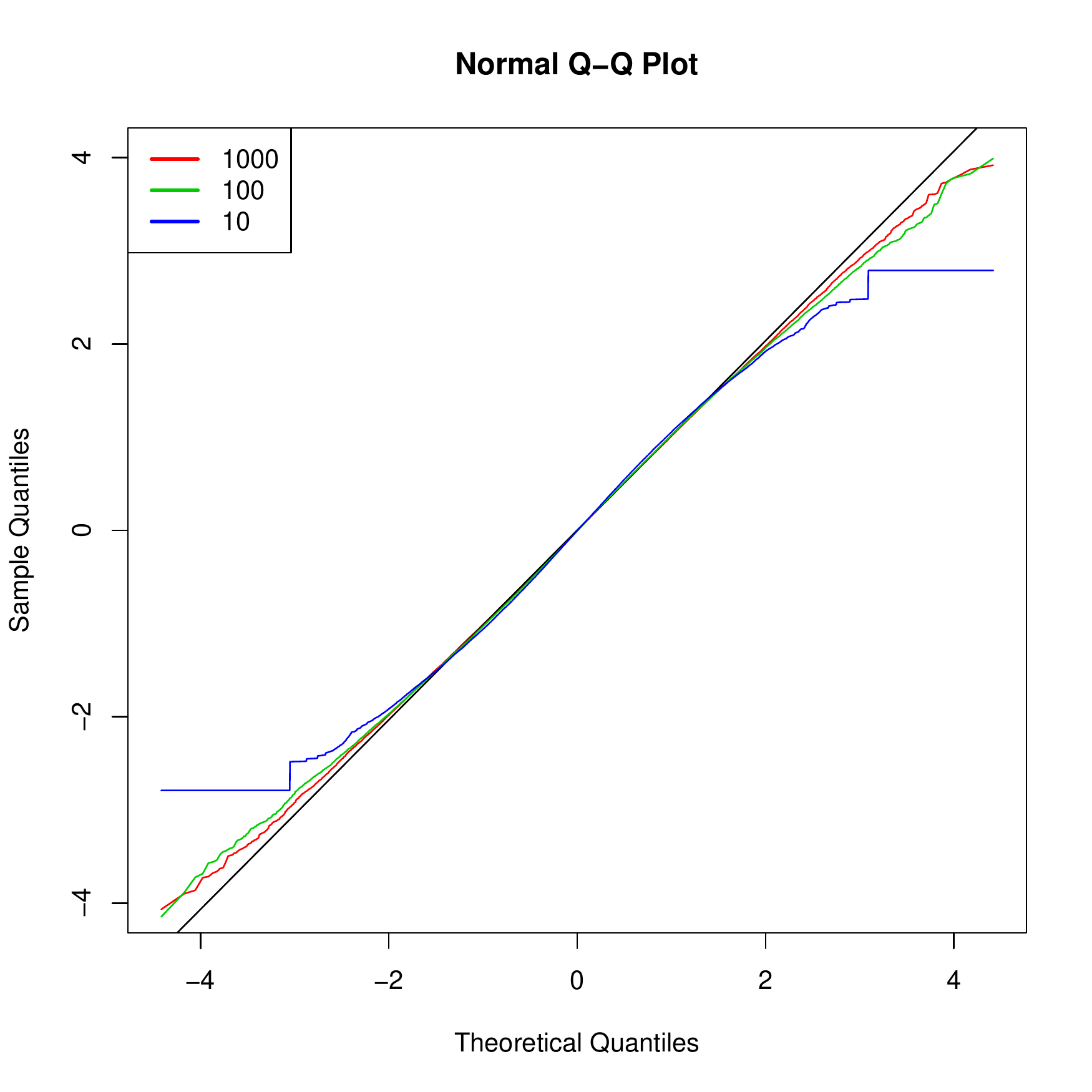}}
\caption{QQ-plot showing the convergence of the single-sample r-statistic $R_0$  to a Gaussian variable  as a function of $N$ \label{fig:qqplot}}
\end{figure}

The distribution of  r-statistic $r=R_0/\sigma_N$ converges slowly to a Gaussian, as illustrated by the qq-plot of Fig.\ \ref{fig:qqplot}.

\section{ROC curves}\label{sec:ROC_curves}

\subsection{One sample}

\begin{figure}
\centerline{\includegraphics[width=0.5\textwidth]{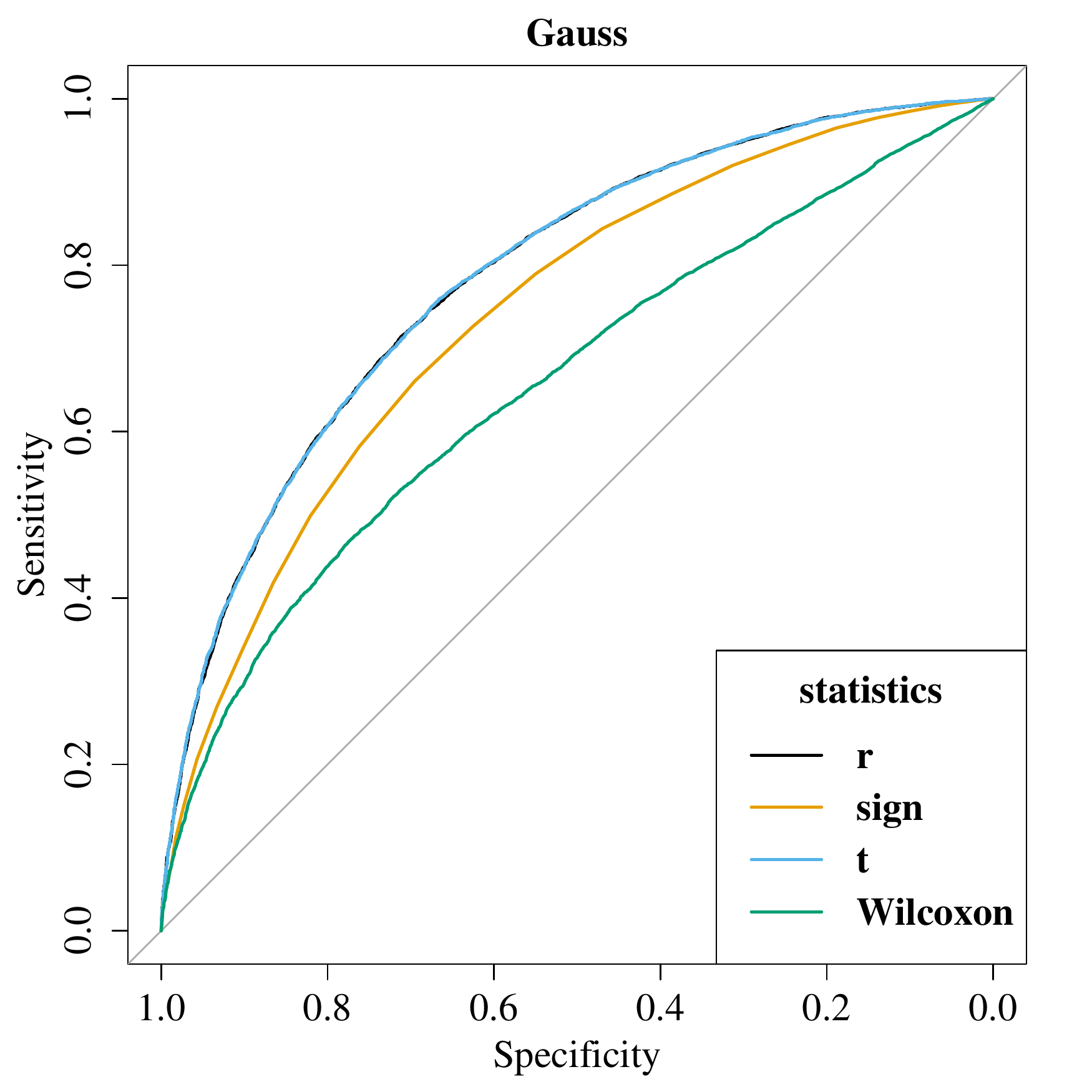}\includegraphics[width=0.5\textwidth]{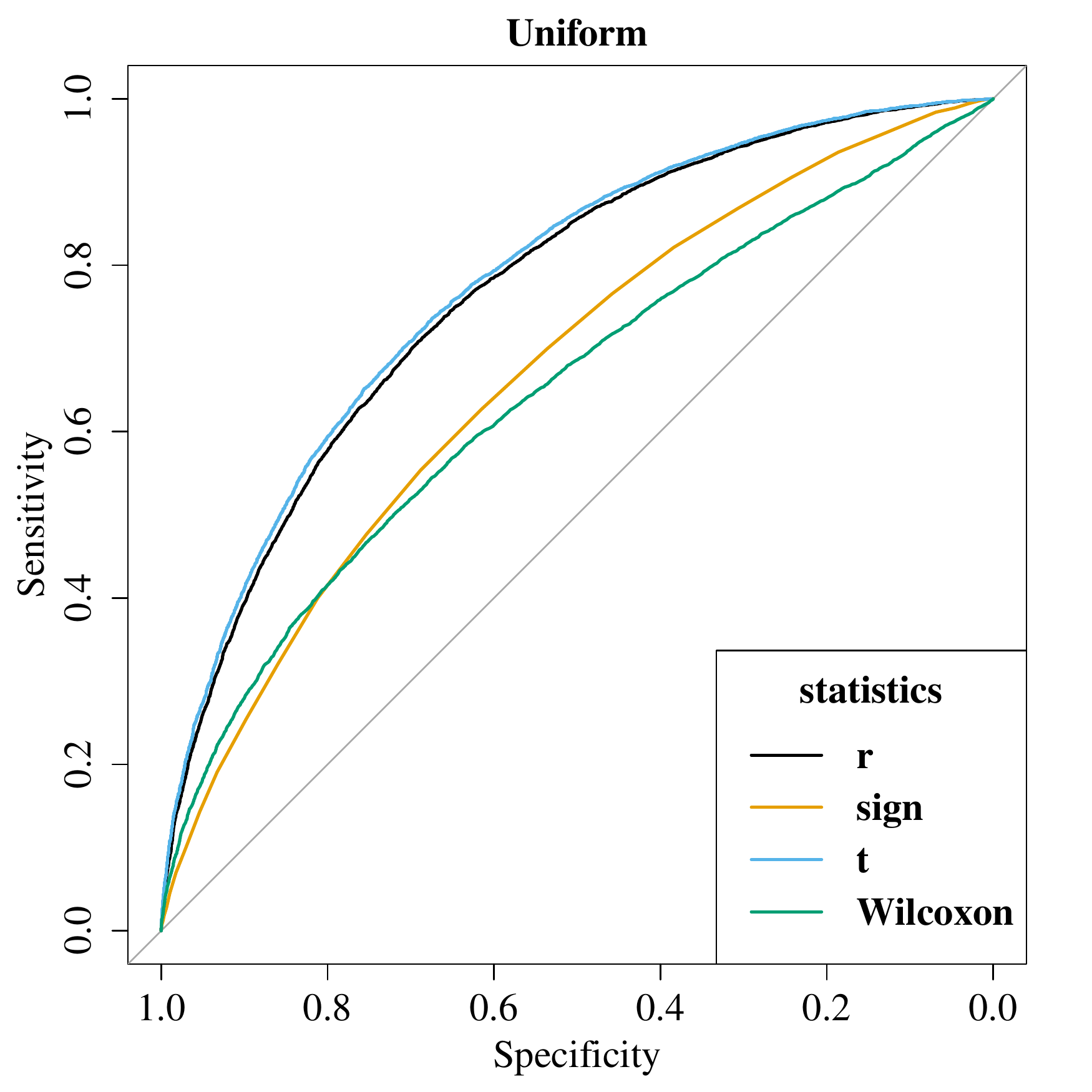}}
\centerline{\includegraphics[width=0.5\textwidth]{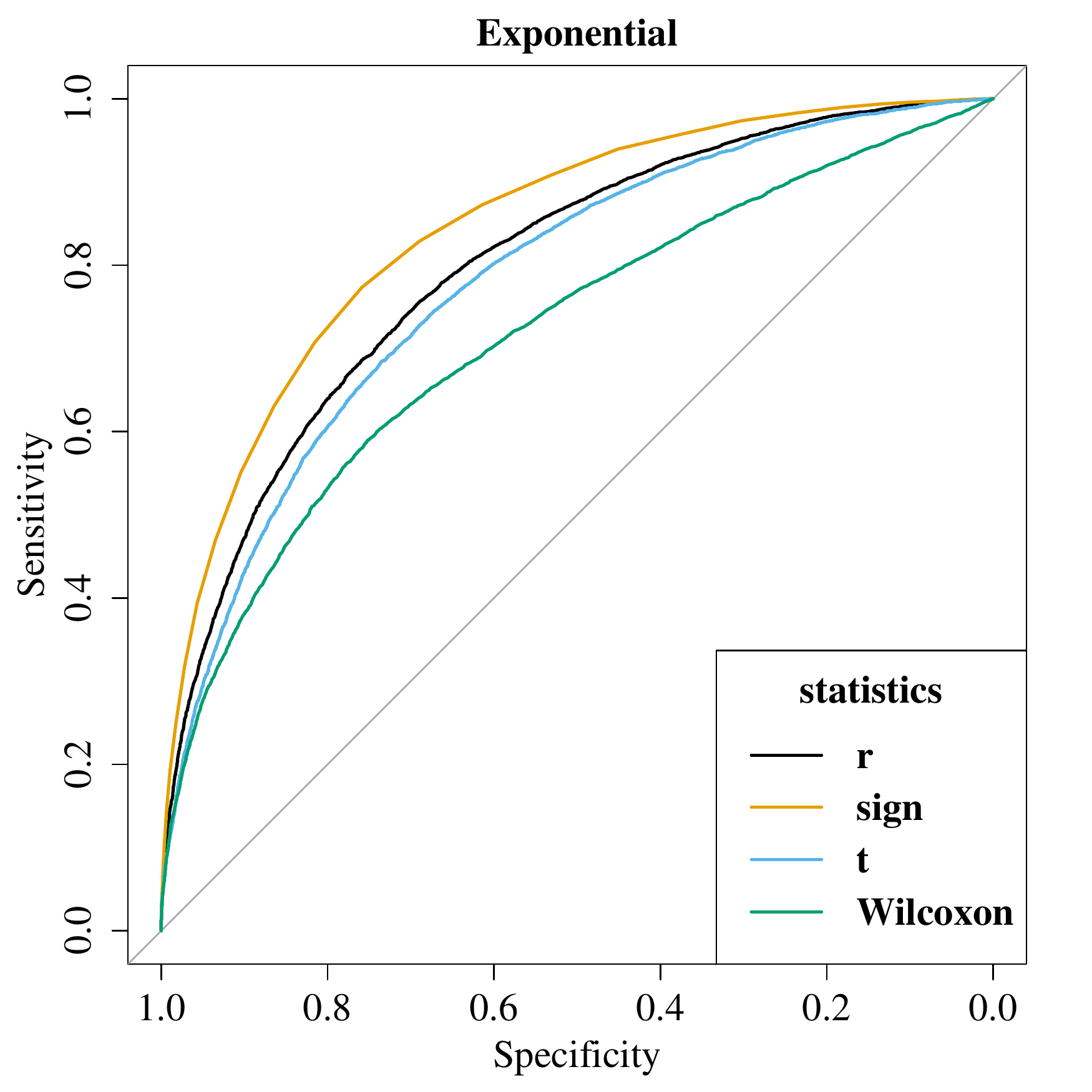}\includegraphics[width=0.5\textwidth]{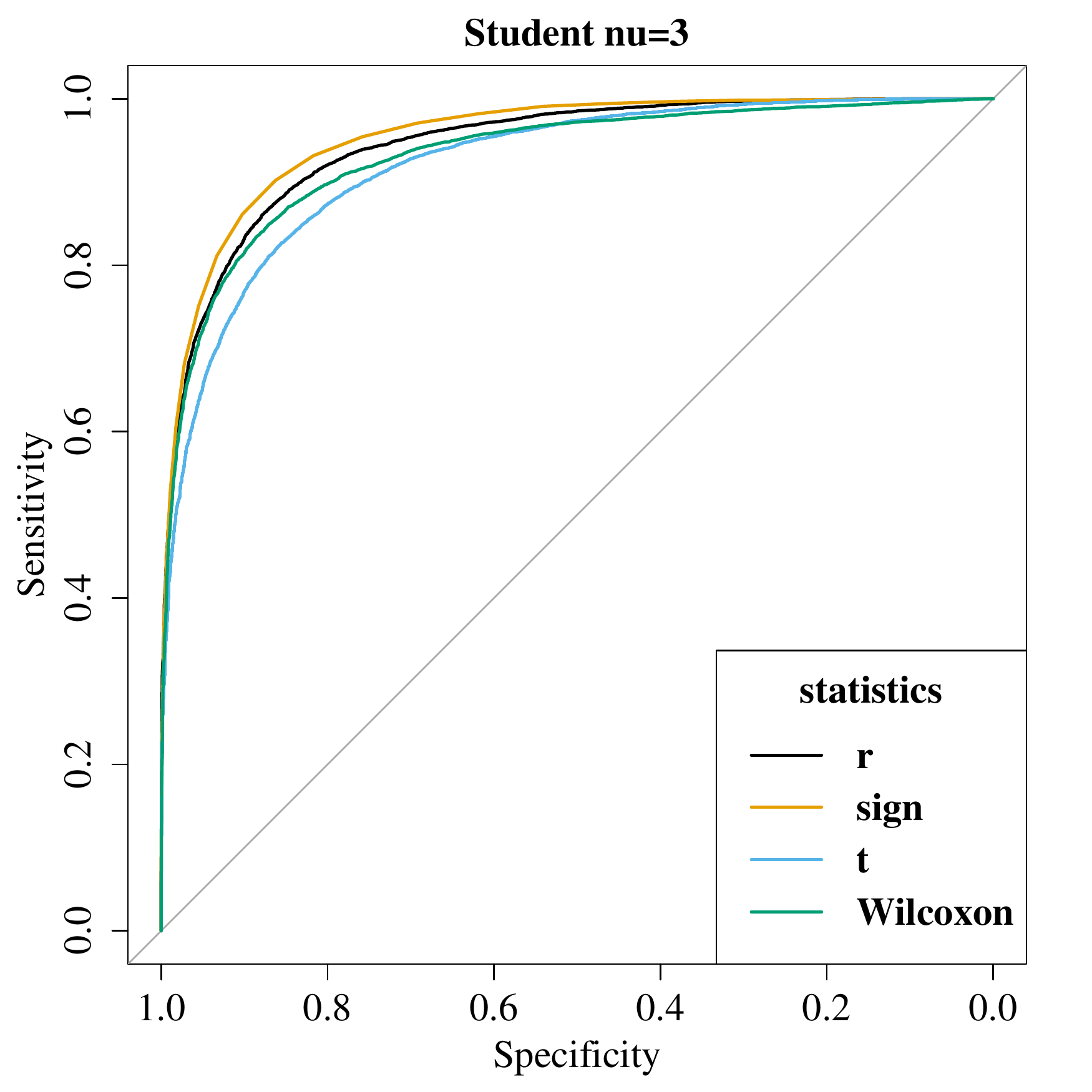}}
\caption{Single-sample situation: ROC curves
for various distributions of $\{x_{n}\}$; the alternative has a signal-to-noise ratio $\theta=\mu/\sigma=0.11$,
10000 samples of length $N=100$, 10000 random permutations per sample. \label{fig:ROCS-single}}

\end{figure}

Figure \ref{fig:ROCS-single} plots ROC curves for the four distributions investigated here. It should be noted that r-statistic curves do not cross those of the other statistics.

\subsection{Two samples}

\begin{figure}
\centerline{\includegraphics[width=0.5\textwidth]{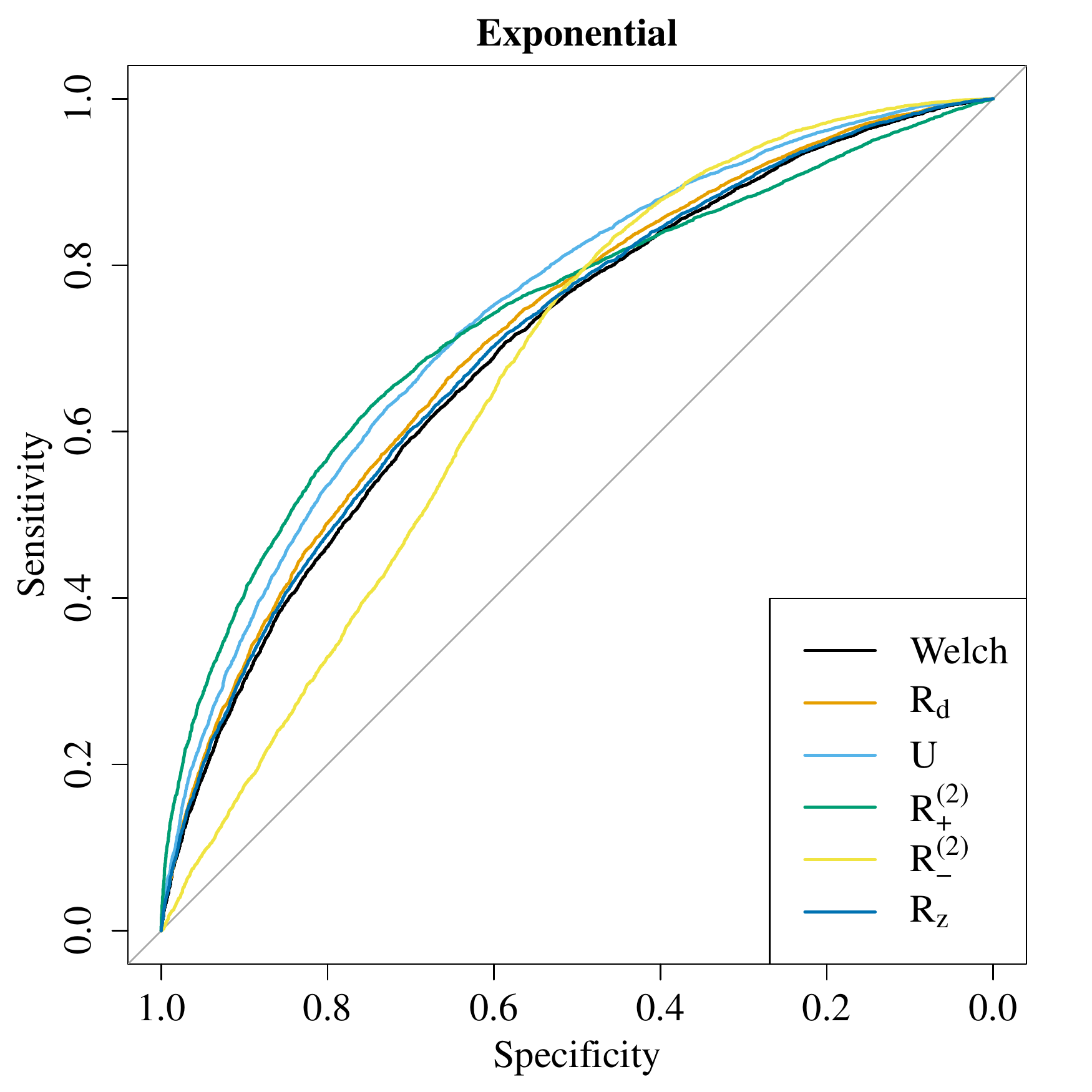}\includegraphics[width=0.5\textwidth]{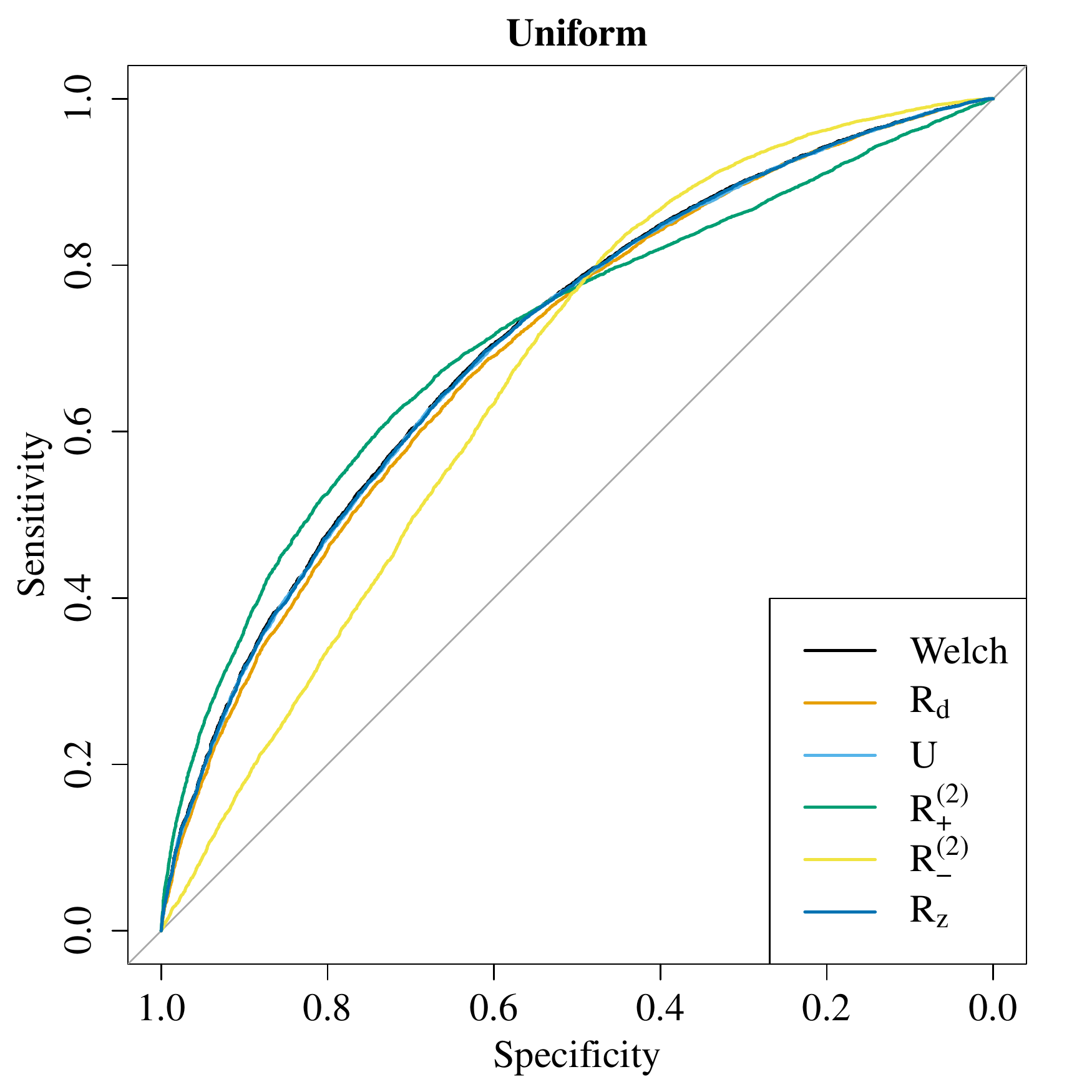}}

\caption{Two-sample situation: ROC curves for various statistics. Both samples have $\sigma=1$, while $E(x)=0$ and $E(y)=1$. $N=100$, 10000
samples per point, 10000 permutations per sample. \label{fig:ROCS-twoappendix}}

\end{figure}

Figure \ref{fig:ROCS-twoappendix} plots ROC curves for the uniform and exponential distributions; those for Gaussian and Student's t distributions are reported in Fig.\ \ref{fig:ROC_2sample}.

\bibliographystyle{rspublicnatwithsort}
\bibliography{biblio}

\end{document}